\documentclass[aps,prx,reprint,longbibliography,amsmath,amssymb,
superscriptaddress,nofootinbib,floatfix]{revtex4-2}
\usepackage{graphicx}
\usepackage{amsthm}
\usepackage{bm}
\usepackage{array}[=2016-10-06]
\usepackage{xcolor}
\usepackage{xurl}
\usepackage{hyperref}
\usepackage{booktabs} 
\hypersetup{colorlinks=true,linkcolor=blue,citecolor=blue,urlcolor=blue,
pdftitle={Monotone Crossing Lattices for Exact Binary Optimization Encodings on Rydberg Arrays},
pdfauthor={Liyin Hu et al.},
pdfkeywords={Rydberg atom arrays, unit-disk graphs, maximum-weight independent set, binary quadratic optimization}}
\newtheorem{theorem}{Theorem}
\newtheorem{lemma}[theorem]{Lemma}
\newtheorem{proposition}[theorem]{Proposition}

\newcommand{\OPT}{\operatorname{OPT}}
\newcommand{\MWIS}{\mathrm{MWIS}}
\newcommand{\CM}{C_{\mathrm M}}
\newcommand{\CQ}{C_{\mathrm Q}}
\newcommand{\dM}{\delta_{\mathrm M}}
\newcommand{\dQ}{\delta_{\mathrm Q}}
\newcommand{\cD}{\mathcal D_{\mathcal R}}
\newcommand{\cC}{\mathcal C}

\begin{document}
\title{Monotone Crossing Lattices for Exact Binary Optimization Encodings on Rydberg Arrays}
\author{Liyin Hu}
\affiliation{Institute of Computing Technology, Chinese Academy of Sciences}
\affiliation{School of Computer Science and Technology, University of Chinese Academy of Sciences}

\author{Shengminjie Chen}\email{csmj@ict.ac.cn}
\affiliation{Institute of Computing Technology, Chinese Academy of Sciences}
\affiliation{School of Computer Science and Technology, University of Chinese Academy of Sciences}

\author{Ran Miao}
\affiliation{Beijing Zhongke Liangshu Technology Co., Ltd}

\author{Zhiyi Huang}
\affiliation{Beijing Zhongke Liangshu Technology Co., Ltd}

\author{Xiaohan Shan}
\affiliation{Beijing Zhongke Liangshu Technology Co., Ltd}

\author{Jialin Zhang}
\affiliation{Institute of Computing Technology, Chinese Academy of Sciences}
\affiliation{School of Computer Science and Technology, University of Chinese Academy of Sciences}

\author{Xiaoming Sun}\email{sunxiaoming@ict.ac.cn}
\affiliation{Institute of Computing Technology, Chinese Academy of Sciences}
\affiliation{School of Computer Science and Technology, University of Chinese Academy of Sciences}

\date{November 25, 2026}

\begin{abstract}
Programmable Rydberg-atom arrays realize maximum-weight independent set encoding on unit-disk graphs under the ideal blockade model, but embedding nonlocal interactions requires auxiliary atoms and limits the size of tractable instances. We introduce a four-state unit-disk crossing cell that couples directly to adjacent cells, with interfaces enforcing one-sided propagation: along occupied paths directed away from the diagonal, copy bits are nondecreasing. For the maximum-weight independent set problem, uniform cell weights guarantee that every ground state of the encoded instance decodes to an optimal solution of the original problem, even when local copies disagree. For binary quadratic objectives subject to pairwise Boolean constraints, every feasible optimum yields consistent active copies and a target weight given by an explicit constant offset minus the source objective. A rectangular variant of the architecture further encodes minimum set cover problem. Prior to constraint-induced site deletions, the full layout requires $2n^2 - n$ sites, halving the leading-order site count of a fixed complete-grid reference, while activity pruning yields an exact site count set by the propagation span. Across 18 paired benchmarks with independent ordering from a shared candidate pool, the layout reduces site counts by $12.1\%\text{–}54.0\%$.  
\end{abstract}
\maketitle

\section{Introduction}
\label{sec:introduction}

Recent experimental advances in the preparation, coherent control, and dynamic reconfiguration of neutral-atom arrays have established Rydberg-atom arrays as a versatile platform for both quantum simulation~\cite{Ebadi2021} and quantum computing~\cite{Bluvstein2024}. Beyond general quantum information processing, programmable Rydberg-atom arrays provide a natural platform for encoding and solving graph-based combinatorial optimization problems~\cite{Pichler2018} by exploiting the Rydberg blockade effect, which suppresses simultaneous Rydberg excitations of nearby atoms and thereby induces effective pairwise exclusion constraints.

Under the idealized blockade approximation, the atomic array can be mapped onto a graph $G=(V,E)$, where each atom corresponds to a vertex and an edge connects two atoms whose separation is smaller than the blockade radius. Consequently, allowed Rydberg excitation configurations correspond to independent sets of the resulting graph. For atoms arranged in a two-dimensional plane, this construction naturally leads to a unit-disk graph~\cite{Pichler2018}. Furthermore, site-dependent detunings provide tunable vertex weights by controlling the energetic preference for Rydberg excitation. In the zero-drive limit, the ground-state problem thus reduces to finding a maximum-weight independent set (MWIS) of the corresponding graph, provided that residual interactions beyond the blockade radius are neglected~\cite{Pichler2018}. This connection has been experimentally demonstrated in maximum independent set optimization~\cite{Ebadi2022}, as well as in weighted-graph optimization using locally controlled light shifts~\cite{deOliveira2025}.

General interaction graphs, however, cannot always be directly realized as native unit-disk
realizations. Encoding their logical interactions introduces auxiliary
atoms, which consume part of a finite array and limit the number of
logical variables it can accommodate. Reducing this auxiliary-atom overhead is therefore
important for extending Rydberg optimization to larger instances with
nonnative connectivity.

Quantum wires mediate nonlocal interactions through auxiliary
chains~\cite{Qiu2020,Kim2022}. Weighted wires are particularly suited to sparse instances that differ from a native unit-disk graph by a small
number of long-range connections~\cite{deOliveira2026Wires}. Their cost
depends on the number, length, and routing of these connections. When
many edges require nonlocal mediation, as can occur in highly connected
graphs, the required wire network increases the auxiliary-atom count. Mao \emph{et al.} addressed this connectivity challenge by
embedding maximum independent set instances on arbitrary graphs into three-dimensional
Rydberg-atom arrays~\cite{Mao2026}.
In two dimensions, crossing-lattice constructions organize arbitrary connectivity into a
systematic arrangement of variable-copy lines and crossing
gadgets~\cite{Nguyen2023}. A complete lattice contains a crossing for
every variable pair, so the size of each repeated cell and its interfaces
determines the leading quadratic cost.  

Other approaches have explored how to encode logical constraints through local geometric structures in blockade graphs, including satisfiability constructions and planar logic primitives~\cite{Jeong2023SAT,Stastny2023}. Parity-based encodings provide another means of representing constrained objectives and higher-order interactions~\cite{Lanthaler2023}. These developments have broadened the range of logical constraints that can be implemented geometrically. Meanwhile, several embedding-level strategies, including variable ordering, shortening copy lines, and graph simplification, have made progress in reducing the auxiliary-atom overhead of individual instances~\cite{Nguyen2023,Schuetz2025}. Although existing strategies can improve the embedding layout and reduce the lengths of individual copy lines, they do not fundamentally eliminate the overhead associated with the repeated crossing structure itself. Building on this idea, we investigate how to compress a crossing lattice while preserving the exact decodability of all optimal solutions of the target problem. 

Our construction uses four sites for the four assignments to a pair of Boolean variables. Adjacent cells interact directly at their boundaries through a one-sided implication. Along a consecutively occupied path directed away from a variable's diagonal site, a local copy can change from $0$ to $1$, and the geometry forbids a change from $1$ to $0$. This rule removes the need for separate connector sites at the interface.

\begin{figure}[b]
\centering
\includegraphics[width=\columnwidth]{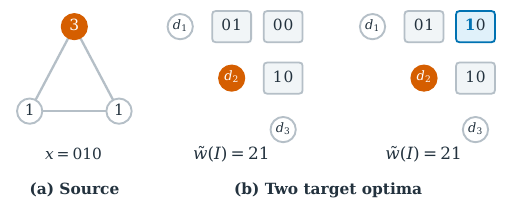}
\caption{Correct maximum-weight independent-set (MWIS) readout
with unequal local copies.
(a) The source triangle has weights $(w_1,w_2,w_3)=(1,3,1)$
and unique optimal incidence vector $x=(x_1,x_2,x_3)=(0,1,0)$.
(b) In the target lattice, the diagonal site $d_i$ carries weight
$w_i$ and reads out $x_i=1$ when selected. Each off-diagonal cell
$(i,j)$ represents the row copy of $x_i$ followed by the column copy
of $x_j$. All three pairs are source edges, so state $11$ is deleted
in every cell. Each retained state has reward
$\delta_{\mathrm M}=\sum_i w_i+1=6$. Both illustrated independent
sets select $d_2$ and decode to $010$. In cell $(1,3)$ the left set
selects $00$, and the right set selects $10$ after the row copy of
$x_1=0$ rises to one. Each set $I$ has target weight
$\tilde{w}(I)=3\times6+3=21$.}
\label{fig:mwis-intro}
\end{figure}

For MWIS, the source rewards reside on the diagonal. In a filled
layout, a diagonal one remains one along its path. Thus two selected
endpoints of a source edge would require state $11$ at their crossing,
which the construction deletes. A copy of a diagonal zero can still
rise to one. Figure~\ref{fig:mwis-intro} shows two optimal target
configurations that differ in this way. Uniform cell rewards ensure
that every target optimum is filled, and its diagonal readout
has maximum source weight.

In a binary quadratic objective, the interaction contribution is evaluated on the crossing copies themselves. Endpoint compensation makes the one-sided implication sufficient for equality at optimality. A diagonal reward cancels an endpoint deduction when the copies agree, and a rise from zero to one incurs a deduction that dominates any possible gain in the source objective. For feasible inputs with allowed-assignment relations on pairs, every target optimum has consistent copies along its active paths and decodes to a constrained source minimizer. A rectangular construction for minimum set cover further illustrates how this implication can express a covering condition. This construction complements geometric encodings of satisfiability, planar logic, and parity constraints~\cite{Jeong2023SAT,Stastny2023,Lanthaler2023}.

The complete triangular layout uses $2n^2-n$ sites for $n$
variables before relation-induced state deletion, compared with
$4n^2-2n-4$ in the fixed complete-grid reference~\cite{UnitDiskMappingSnapshot}.
Activity pruning reduces the count according to the propagation span.
We quantify site savings using both a shared order and separate
order selection from a common candidate set. 

Section~\ref{model} defines the optimization and blockade models.
Sections~\ref{sec:mwis} and \ref{sec:extensions} prove the MWIS and
quadratic encoding guarantees. Section~\ref{sec:setcover} gives the
rectangular set-cover construction, and Sec.~\ref{sec:resources}
analyzes the site count.

\section{Optimization problems and the ideal blockade model}
\label{model}

\subsection{Source problems and target graphs}

\paragraph{Source problems.}
For a positive integer $k$, write $[k]=\{1,\ldots,k\}$.
Let $G=(V,E)$ be a simple graph with vertex set $V$ and edge set $E$.
An independent set of $G$ is a set of vertices
containing no adjacent pair. Given vertex weights $w_i$, the
 maximum-weight independent set (MWIS) problem asks for
\begin{equation*}
\OPT_{\MWIS}(G,w)
=
\max_{S\subseteq V:\,S\text{ independent}}
\sum_{i\in S}w_i.
\end{equation*}
We consider simple graphs with rational vertex weights. After preprocessing, we assume without loss of
generality that the residual graph has positive weights and no isolated vertices. Relabel a nonempty residual graph by $V=[n]$. Throughout, $n$ denotes
the number of residual variables unless otherwise specified.

For constrained binary quadratic optimization, we consider pairwise
Boolean relations in a binary quadratic program~\cite{Bayani2024BQP}. Given rational coefficients
$c$, $h_i$, and $J_{ij}$, the problem is $\min_{x\in\cD}Q(x)$ for binary variables $x=(x_1,\ldots,x_n)\in\{0,1\}^n$, with
\begin{equation}
Q(x)
=
c+\sum_{i=1}^n h_i x_i
+\sum_{1\le i<j\le n}J_{ij}x_i x_j.
\label{eq:source-qubo}
\end{equation}
For each constrained pair $(i,j)\in\mathcal P$, with $i<j$,
let $R_{ij}\subseteq\{0,1\}^2$ specify the allowed assignments
to $(x_i,x_j)$. The feasible domain is
\begin{equation}
\cD
=
\bigl\{
x\in\{0,1\}^n:
(x_i,x_j)\in R_{ij}
\text{ for every }(i,j)\in\mathcal P
\bigr\}.
\label{eq:domain}
\end{equation} 
An explicitly empty $R_{ij}$ certifies infeasibility before construction.
The quadratic encoding below assumes each specified relation is nonempty; the relations can still be jointly infeasible.
Setting $\mathcal P=\varnothing$ recovers the quadratic unconstrained binary optimization problem (QUBO)~\cite{Glover2019QUBO,Lucas2014}.
 
\paragraph{Target graphs.} A graph $H=(V(H),E(H))$ is a unit-disk graph (UDG) if its vertices admit
distinct planar positions $q(v)\in\mathbb R^2$ satisfying, for distinct $u,v\in V(H)$,
\begin{equation*}
\{u,v\}\in E(H)
\quad\Longleftrightarrow\quad
\|q(u)-q(v)\|_2\le1. 
\end{equation*}
The target problem is maximum-weight independent set on a unit-disk graph (UDG-MWIS), with positive rational weights $\omega_v$ for $v\in V(H)$ and optimal value $\OPT_{\MWIS}(H,\omega)$.

Each exact encoding specifies the target graph, its coordinates and weights, an explicit relation between the source and target optimal values, and a decoding map. For feasible source instances, this map recovers a source optimum from every target optimum.

\subsection{Rydberg realization of the target graph}
\begin{figure*}[t]
\centering
\includegraphics[width=\textwidth]{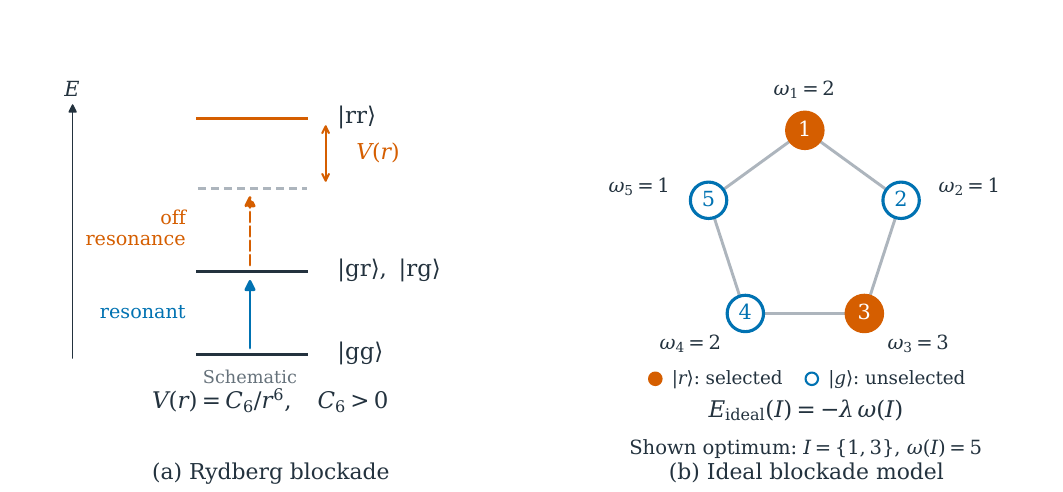}
\caption{Rydberg blockade and its graph interpretation.
(a) Schematic laboratory-frame two-atom levels at single-atom resonance.
The dashed level marks the unshifted $|rr\rangle$ energy. The repulsive
interaction shifts this level by $V(r)=C_6/r^6$, moving the second
excitation out of resonance.
(b) In the ideal blockade model, atoms are vertices and pairs within the
chosen cutoff $R_b$ define edges. The atoms shown lie on a regular pentagon
of radius $0.75R_b$. Orange sites mark the excited atoms;
the labels $\omega_v$ give vertex rewards assigned through
$\Delta_v=\lambda\omega_v$, with $\lambda>0$.
The highlighted sites form the unique maximum-weight independent set of
the illustrated graph, with total weight $5$ and zero-drive ideal energy
$-5\lambda$.}
\label{fig:blockade-overview}
\end{figure*}

For atoms at positions $\mathbf r_v\in\mathbb R^2$, the two-level
Rydberg Hamiltonian at time~$\tau$ is~\cite{Pichler2018,Nguyen2023}
\begin{equation*}
\begin{aligned}
\widehat H_{\mathrm{Ryd}}(\tau)
&=\sum_v\frac{\Omega_v(\tau)}2\sigma_v^x-\sum_v\Delta_v(\tau)n_v\\
&\quad+\sum_{\{u,v\}}\frac{C_6}{\|\mathbf r_u-\mathbf r_v\|_2^6}n_un_v,
\end{aligned} 
\end{equation*}
with $n_v=|r\rangle_v\langle r|$, $\sigma_v^x=|g\rangle_v\langle r|+|r\rangle_v\langle g|$, and $C_6>0$. Here $\Omega_v(\tau)$ is the Rabi frequency of the transverse drive and $\Delta_v(\tau)$ is the local detuning; $\hbar=1$.

In the ideal blockade model, simultaneous excitations of atoms associated with adjacent vertices are strictly forbidden, while interactions between nonadjacent vertices are neglected. To realize a target UDG $H$, we place the atoms at positions $\mathbf r_v=R_bq(v)$, where $R_b$ is the distance cutoff used in the ideal model. Under zero drive, i.e., $\Omega_v=0$, and with local detunings set to $\Delta_v=\lambda\omega_v$ for some $\lambda>0$, the allowed excitation configurations correspond exactly to the independent sets of $H$. In particular, the energy of an allowed configuration $I$ is 
\begin{equation*}
E_{\mathrm{ideal}}(I)=-\lambda\omega(I), 
\end{equation*}
where $\omega(I)=\sum_{v\in I}\omega_v$. Consequently, the ground-state configurations of the ideal blockade model are exactly the maximum-weight independent sets of the target graph.

\paragraph{Finite interactions.}
Figure~\ref{fig:blockade-overview} illustrates the ideal correspondence.
The geometric separation in Lemma~\ref{lem:geometry} guarantees the
same cutoff graph if each atom is displaced by less than $\eta R_b/2$.
For the full Hamiltonian, nonedge interactions and finite edge
penalties affect the energy ordering at zero drive. Define the
dimensionless detuning ratio $t=\lambda/(C_6/R_b^6)$. In the ten-site quadratic
example of Appendix~\ref{app:examples}, two competitors rule out the
graph optimum as a zero-drive ground state at every $t>0$ under uniform rescaling $\Delta_v=\lambda\omega_v$. An exact enumeration further
rules out every configuration with the optimal diagonal readout at
these fixed coordinates and relative detunings. The competing energies
and the enumeration are given in Appendix~\ref{app:finite-interaction}.
 
\section{Monotone crossing lattice and MWIS encoding}
\label{sec:mwis} 

The lattice encodes each Boolean variable at a diagonal site and places its copies along a row and a column. The same geometry supports a complete layout and a layout pruned to the required interactions.

\subsection{Four-state cells with monotone interfaces}\label{sec:four-state-geometry}

The complete upper-triangular layout contains one diagonal
vertex $d_i$ at each diagonal index $(i,i)$ for $i=1,\ldots,n$, and one four-state cell
$(i,j)$ for every $1\le i<j\le n$. Before pruning, each off-diagonal
cell contains four vertices $v_{ij}^{ab}$, with $a,b\in\{0,1\}$
representing copies of $x_i,x_j$, respectively. At $(i,i)$ only
$v_{ii}^{11}=d_i$ is present. The diagonal bit $x_i$ is the indicator
that $d_i$ belongs to the independent set.

Place the state vertices at the dimensionless coordinates
\begin{equation*}
q(v_{ij}^{ab})=
\left(\frac{27(j-1)}{20}+\frac{3a}{5},
      -\frac{27(i-1)}{20}+\frac{3b}{5}\right).
\end{equation*}
The same formula gives $q(d_i)$ with $j=i$ and $a=b=1$.

\begin{lemma}[Four-state unit-disk geometry]
\label{lem:geometry}
The coordinates above realize the complete layout with edge lengths at most
$1-\eta$ and nonedge distances at least $1+\eta$ for a
constant $\eta>0$ independent of $n$. The vertices of each cell form a clique, and the intercell edges join horizontally or vertically adjacent cells as follows:
\begin{align}
\{v_{ij}^{ab},v_{i,j+1}^{a'b'}\}\in E(H)
&\iff (a,a')=(1,0),
\label{eq:horizontal}\\
\{v_{ij}^{ab},v_{i+1,j}^{a'b'}\}\in E(H)
&\iff (b,b')=(0,1).
\label{eq:vertical}
\end{align}
Equation~\eqref{eq:horizontal} applies for $1\le i\le j<n$ and
Eq.~\eqref{eq:vertical} for $1\le i<j\le n$, in each case for
pairs of vertices present in the indicated cells. At $i=j$ the
horizontal rule joins $d_i$ to right-cell states with $a'=0$.
At $j=i+1$ the vertical rule joins upper-cell states with $b=0$
to $d_j$.
\end{lemma}

\begin{figure*}[t]
\centering
\includegraphics[width=\textwidth]{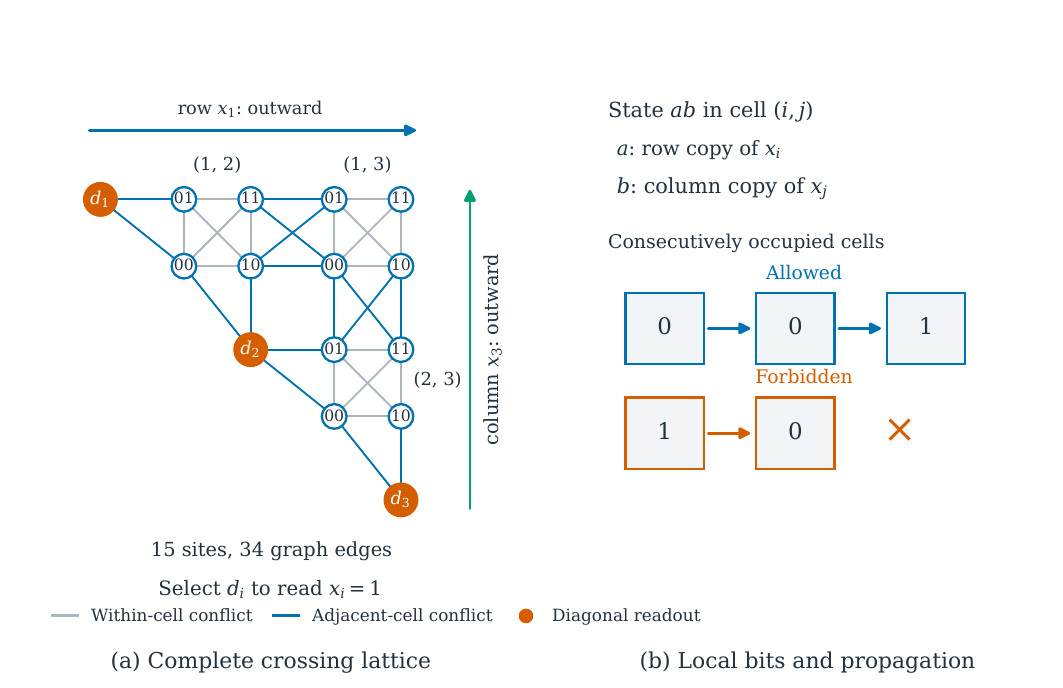}
\caption{The crossing lattice and its local implication rule.
(a) The complete three-variable realization has 15 sites and 34 edges.
Labels $ab$ identify the row and column copies in each cell. Gray and
blue edges denote within-cell and adjacent-cell conflicts; orange sites
provide diagonal readout. Arrows orient propagation away from the diagonal.
(b) Along consecutively occupied cells, the geometry permits $0\to1$
and excludes $1\to0$. Uniform cell rewards give optimal MWIS readout,
while endpoint compensation provides the quadratic encoding.
}
\label{fig:geometry}
\end{figure*}

Appendix~\ref{app:geometry} proves the distance bounds.
Orient each path away from $d_i$, rightward along the row and
upward along the column. Across consecutive occupied cells the
conflict edges forbid $1\to0$, so the copy bits are nondecreasing
in the outward direction. If the first cell is occupied, its
copy bit is at least $x_i$: a selected $d_i$ blocks a zero, and
$x_i=0$ permits either value. Figure~\ref{fig:geometry}
illustrates these interfaces.

\subsection{Complete and activity-pruned layouts}
Let $F\subseteq\{(i,j):1\le i<j\le n\}$ be the required pairs.
For MWIS and constrained quadratic optimization, respectively, use
\begin{align*}
F&=\{(i,j):1\le i<j\le n,\ \{i,j\}\in E(G)\},\\
F&=\{(i,j):1\le i<j\le n,\ J_{ij}\ne0\}\cup\mathcal P.
\end{align*}
In the complete layout set $R_i=n$ and $L_i=1$ for every $i\in[n]$.
In the activity-pruned layout set, for each $i\in[n]$,
\begin{align*}
R_i&=\max\bigl(\{i\}\cup\{j:(i,j)\in F\}\bigr),\\
L_i&=\min\bigl(\{i\}\cup\{k:(k,i)\in F\}\bigr).
\end{align*}
In either layout, $R_i$ is the terminal column of the horizontal
segment and $L_i$ the terminal row of the vertical segment for $x_i$.
The horizontal segment is nonempty if $R_i>i$ and the vertical segment
is nonempty if $L_i<i$.

For each off-diagonal cell $(i,j)$, let $\epsilon_{ij}^{h}=1$ when $j\le R_i$ and
$\epsilon_{ij}^{v}=1$ when $i\ge L_j$, with each flag equal to zero otherwise. An active bit retains both values, while an inactive bit is fixed to one. Along either outward path, inactive bits occur only beyond the last active copy. Fixing them to one lets either active bit value reach the outer
boundary without a forbidden $1\to0$ transition. The set
$\mathcal A_{ij}\subseteq\{0,1\}^2$ retained before relation deletion is
\begin{equation}
\mathcal A_{ij}
=
\begin{cases}
\{00,01,10,11\}, & (\epsilon_{ij}^{h},\epsilon_{ij}^{v})=(1,1),\\
\{01,11\},      & (\epsilon_{ij}^{h},\epsilon_{ij}^{v})=(1,0),\\
\{10,11\},      & (\epsilon_{ij}^{h},\epsilon_{ij}^{v})=(0,1),\\
\emptyset,     & (\epsilon_{ij}^{h},\epsilon_{ij}^{v})=(0,0).
\end{cases}
\label{eq:states}
\end{equation}

We retain $\{v_{ij}^{ab}:ab\in\mathcal A_{ij}\}$ at this stage.
At each required pair $(i,j)\in F$, both bits are active, so activity pruning leaves all four assignments available for the subsequent relation deletion. For MWIS, delete the state $11$ whenever $\{i,j\}\in E(G)$. For the pairwise constraints in Eq.~\eqref{eq:domain}, retain only states belonging to the allowed relation $R_{ij}$ at each constrained pair. All retained vertices keep their original coordinates, so the resulting graph is an induced unit-disk subgraph of the complete realization.
 
A target independent set is \emph{filled} if it selects exactly one vertex from every retained off-diagonal cell.
For each feasible source assignment $x$, define its \emph{canonical encoding} $I_x$ by selecting $d_i$ if and only if $x_i=1$ and selecting, in each retained cell $(i,j)$, the state $v_{ij}^{a_{ij}b_{ij}}$, where
\begin{equation*}
a_{ij}
=
\begin{cases}
x_i,&\epsilon_{ij}^{h}=1,\\
1,&\epsilon_{ij}^{h}=0,
\end{cases}
\qquad
b_{ij}
=
\begin{cases}
x_j,&\epsilon_{ij}^{v}=1,\\
1,&\epsilon_{ij}^{v}=0.
\end{cases} 
\end{equation*}
These choices give a filled independent set by the boundary
argument in Appendix~\ref{app:pruning}. More generally, call a
filled independent set \emph{canonical} if every active copy agrees
with its diagonal bit. All copies are active in the complete layout.

\subsection{Exact MWIS encoding with uniform cell rewards}
The diagonal readout of a target independent set $I$ is
$S(I)=\{i\in[n]:d_i\in I\}$. Its incidence vector is
$x(I)\in\{0,1\}^n$, with $x_i(I)=1$ exactly when $i\in S(I)$.

Let $\cC$ contain the off-diagonal cells of the chosen layout before relation-induced state deletion, and let $K=|\cC|$. Thus $K=\binom n2$ for the complete layout, and $K=K_1+K_2$ for the pruned layout, where $K_s$ counts cells with $s$ active bits. We retain source weights on the diagonal and give all retained state vertices a uniform reward.

\begin{equation*}
\begin{aligned}
W&=\sum_i w_i,
\qquad
\dM=W+1,
\qquad
\CM=K\dM,\\
\widetilde w(d_i)&=w_i,
\qquad
\widetilde w(v_{ij}^{ab})=\dM.
\end{aligned}
\end{equation*}

\begin{theorem}[Exact MWIS encoding with diagonal readout]
\label{thm:wmis}
Use either layout, take $F=\{(i,j):i<j,\ \{i,j\}\in E(G)\}$,
and delete state $11$ at every pair in $F$. Then
\begin{equation}
\OPT_{\MWIS}(H,\widetilde w) = \CM+\OPT_{\MWIS}(G,w).
\label{eq:wmis-opt}
\end{equation}
Every maximum-weight independent set $I$ of $H$ is filled, and its diagonal readout $S(I)$ is a maximum-weight independent set of $G$.
\end{theorem}

\begin{proof}
Write $w(S)=\sum_{i\in S}w_i$. Each source independent set $S$
has a canonical encoding of weight $\CM+w(S)$. Taking the
maximum, with the empty source set as a feasible choice, gives
\begin{equation*}
\OPT_{\MWIS}(H,\widetilde w)
\ge\CM+\OPT_{\MWIS}(G,w)\ge\CM.
\end{equation*} 

A target independent set with an empty cell has at most $K-1$
cell vertices and diagonal reward at most $W$. Its weight is
therefore at most
\begin{equation*}
(K-1)\dM+W=\CM-1.
\end{equation*}
Therefore, every target optimum is filled.

Let $I$ be filled and write $x=x(I)$. For each retained cell
$(i,j)$, write $v_{ij}^{a_{ij}(I)b_{ij}(I)}$ for its unique
selected vertex.
At each source edge $\{i,j\}$, both local bits are active, and all
cells along their paths from the diagonal are occupied.
Lemma~\ref{lem:geometry} gives $x_i\le a_{ij}(I)$ and $x_j\le b_{ij}(I)$.
Deletion of state $11$ then implies
$x_ix_j\le a_{ij}(I)b_{ij}(I)=0$, so $S(I)$ is independent in $G$.
Its target weight is
\begin{equation*}
\widetilde w(I)=\CM+w(S(I))
\le\CM+\OPT_{\MWIS}(G,w). 
\end{equation*}
The bounds coincide, and equality requires $S(I)$ to attain the
source optimum.
\end{proof}

The canonical encoding gives one filled representation for each source independent set. The triangle in Fig.~\ref{fig:mwis-intro} has three optimal target configurations, with states $00$, $01$, or $10$ in cell $(1,3)$ and identical diagonal readout.

\section{Exact encoding of constrained quadratic objectives}
\label{sec:extensions}

The quadratic term $-3x_1x_2$ in $Q(x)=x_1+x_2-3x_1x_2$ illustrates why diagonal readout alone is insufficient. Suppose the linear terms were evaluated at the diagonal and the product on the crossing state, with only the monotone interface enforcing consistency. Diagonal readout $00$ could be paired with the crossing state $11$, producing a spurious objective value of $-3$ even though the true minimum is $-1$. We prevent this discrepancy by charging each outward rise through its terminal bit.

\subsection{Endpoint weights}
A variable absent from all nonzero quadratic terms and all prescribed
relations can be minimized independently. Add $\min\{0,h_i\}$ to $c$
and record the minimizing bit for reconstruction. If this eliminates
all variables, the recorded bits solve the instance directly.

\begin{figure*}[tbp]
\centering
\includegraphics[width=\textwidth]{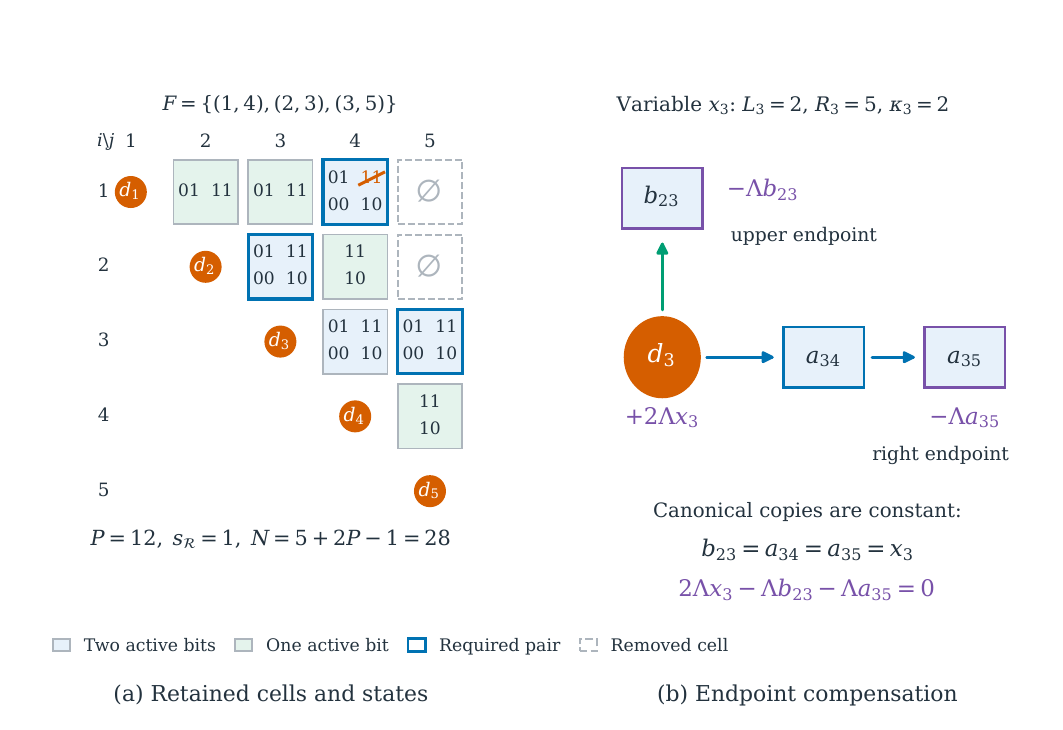}
\caption{Activity pruning and endpoint compensation.
(a) Retained states for $F=\{(1,4),(2,3),(3,5)\}$, with mutual
exclusion at $(1,4)$. Exterior row and column indices identify the cells.
Cell $(3,4)$ remains for propagation although it is not in $F$.
The retained layout has $P=12$, $s_{\mathcal R}=1$, and 28 vertices.
(b) The active paths of $x_3$ end at $(2,3)$ and $(3,5)$.
Their endpoint deductions cancel the diagonal compensation when the
local copies equal $x_3$.}
\label{fig:pruning-compensation}
\end{figure*}

Use the terminal indices of the chosen layout from
Sec.~\ref{sec:mwis}. Variable elimination ensures that every
residual variable occurs in a pair in $F$. Thus $L_i<i$ or
$R_i>i$, and the number of nonempty segments is
\[
\kappa_i=\mathbf 1_{L_i<i}+\mathbf 1_{R_i>i}\in\{1,2\},
\]
where $\mathbf 1$ is the indicator of the stated condition.
Define the coefficient budgets
\begin{align*}
A_+&=\sum_i\max\{h_i,0\}+\sum_{i<j}\max\{J_{ij},0\},\\
A_-&=\sum_i\max\{-h_i,0\}+\sum_{i<j}\max\{-J_{ij},0\},\\
A&=A_++A_-. 
\end{align*}
Set the weight scales to
\begin{equation}
\Lambda=A+1,\qquad\dQ=3\Lambda,\qquad\CQ=K\dQ,
\label{eq:scales}
\end{equation}
and assign the vertex weights
\begin{align}
\omega(d_i)&=\kappa_i\Lambda-h_i,\label{eq:diagonal-weights}\\
\omega(v_{ij}^{ab})&=\dQ-J_{ij}ab
-\Lambda\bigl(\delta_{j,R_i}a+\delta_{i,L_j}b\bigr).
\label{eq:cell-weights}
\end{align}
Here $\delta_{u,v}$ is the Kronecker delta, equal to one if $u=v$
and zero otherwise. The baseline rewards are $\dM$ and $\dQ$.
If a cell carries no quadratic term, $J_{ij}=0$. The terms $-h_i$
and $-J_{ij}ab$ encode the objective, while $\dQ$ is the base
reward for a filled cell. Each nonempty segment contributes a
diagonal reward $\Lambda x_i$ and a terminal deduction equal to
$\Lambda$ times its terminal bit. These terms cancel on a
canonical segment Fig.~\ref{fig:pruning-compensation}(b).
All target weights satisfy
\begin{equation}
1\le\omega(v)\le4A+3.
\label{eq:weight-range}
\end{equation}

\subsection{Optimality and decoding}
The weight balance has two cases. In a filled layout, each outward
$0\to1$ rise leaves a net deduction $\Lambda$. With empty cells,
a segment can gain up to $\Lambda$ from a missed terminal deduction
or an interrupted path. Each empty cell can account for such a gain
on at most two segments and loses a base reward of $3\Lambda$.
In both cases the net loss of at least $\Lambda$ dominates the range
$A$ of the objective contributions. The following lemma makes this
comparison precise.

\begin{lemma}[Canonical weight identity and separation]
\label{lem:canonical-separation}
For either layout, use its own terminal indices $R_i,L_i$ in
Eqs.~\eqref{eq:scales}--\eqref{eq:cell-weights}. Assume every prescribed
relation is nonempty.
Every feasible assignment $x\in\cD$ has a canonical
independent-set encoding $I_x$ satisfying
\begin{equation}
\omega(I_x)=\CQ+c-Q(x).
\label{eq:canonical-identity}
\end{equation}
Every canonical independent set is $I_x$ for exactly one
$x\in\cD$, given by its diagonal readout.
Moreover, every noncanonical target independent set $I$
satisfies
\begin{equation}
\omega(I)\le\CQ-A_+-1.
\label{eq:noncanonical-bound}
\end{equation}
\end{lemma}

The proof is given in Appendix~\ref{app:quadratic-proof}.

\begin{theorem}[Quadratic optimization with pairwise constraints]
\label{thm:qubo}
Under the assumptions of Lemma~\ref{lem:canonical-separation},
\begin{equation}
\cD\ne\varnothing
\quad\Longleftrightarrow\quad
\OPT_{\MWIS}(H,\omega)\ge\CQ-A_+.
\label{eq:feasibility}
\end{equation}
If feasible, every target optimum is canonical and its
diagonal readout minimizes $Q$ over $\cD$, with
\begin{equation}
\OPT_{\MWIS}(H,\omega)
=
\CQ+c-\min_{x\in\cD}Q(x).
\label{eq:qubo-opt}
\end{equation}
If infeasible,
$\OPT_{\MWIS}(H,\omega)\le\CQ-A_+-1$.
\end{theorem}

\begin{proof}
For any feasible assignment $x\in\cD$,
Eq.~\eqref{eq:canonical-identity} and
$Q(x)-c\le A_+$ give
\[
\omega(I_x)\ge\CQ-A_+.
\]
By Eq.~\eqref{eq:noncanonical-bound}, every
noncanonical independent set has strictly smaller
weight. Thus, whenever $\cD\ne\varnothing$,
every target optimum is canonical.

If $\cD=\varnothing$, no canonical independent set
exists, and Eq.~\eqref{eq:noncanonical-bound} applies
to every target independent set. This proves
Eq.~\eqref{eq:feasibility} and the infeasibility bound.

For feasible instances, maximizing
Eq.~\eqref{eq:canonical-identity} over $\cD$ gives
Eq.~\eqref{eq:qubo-opt}.
Since every target optimum is canonical, its diagonal
readout attains the minimum of $Q$ over $\cD$.
\end{proof}

Any independent set of weight at least $\CQ-A_+$ yields a feasible
diagonal readout. A weight below this threshold certifies infeasibility
when it is the target optimum or an upper bound on it. Appendix~\ref{app:examples}
illustrates the latter case with an optimum of $98$ and a threshold of $99$.
As an independent check before constructing the target graph,
2-SAT can decide whether the pairwise relations have a common assignment:
each forbidden pair $(a,b)$ contributes the clause
$(x_i\ne a)\lor(x_j\ne b)$~\cite{Aspvall1979}.

\paragraph{Additive loss under decoding.}
For a feasible instance, write $W^*=\OPT_{\MWIS}(H,\omega)$ and
$Q^*=\min_{x\in\cD}Q(x)$. Every independent set $I$ with
$W^*-\omega(I)<1$ is canonical, since $W^*\ge \CQ-A_+$ and
Eq.~\eqref{eq:noncanonical-bound} bounds every noncanonical weight.
Its diagonal readout $x(I)$, the incidence vector of $S(I)$, satisfies
\[
Q(x(I))-Q^*=W^*-\omega(I).
\]
The unit separation here uses the normalization in Eq.~\eqref{eq:scales}.

\section{Rectangular encoding of minimum set cover}
\label{sec:setcover}
The set-cover problem requires every element to be contained in at least one selected set. To express this condition, we replace the triangular arrangement of variable pairs with a rectangular array of set-element incidences. For a universe $U=\{u_1,\ldots,u_r\}$ and sets $S_1 \subseteq U,\ldots,S_m \subseteq U$, the minimum set-cover problem is defined by
\begin{equation*}
C^*=\min\{|C|:C\subseteq[m],\ \bigcup_{i\in C}S_i=U\}.
\end{equation*}
For an empty universe, an uncovered element, or at most one
candidate set, decide the instance directly. For the remaining case $m\ge2$, we construct an $m\times r$ rectangular array, with one row for each set and one column for each element.

Each nonincidence cell contains four states $L_0,L_1,R_0,R_1$, where $L_b$ and $R_b$ denote the left and right horizontal types,
respectively, and $b\in\{0,1\}$ is the vertical bit. These port labels apply
only to the rectangular cells.
At an incidence cell, the two right-type states
$R_0,R_1$ are replaced by a single state $T$
positioned midway between them.
The state $T$ retains the right horizontal type and the associated horizontal conflicts
but has no vertical intercell conflicts, allowing
it to interrupt vertical propagation along its column. 
 
In column $j$, let $f_j$ and $\ell_j$ be the first and last incidence
rows. Delete bit-one states at $i\le f_j$ and bit-zero states at
$i\ge\ell_j$, retaining $T$. For any cover $C$, choose a witness $i_*\in C$ with
$u_j\in S_{i_*}$, so $f_j\le i_*\le\ell_j$. Selecting $T$
at $(i_*,j)$ permits ordinary states with bit zero above $i_*$
and bit one below it. Thus the boundary restrictions preserve an encoding of every cover.
A rectangular column is filled when it contains one selected
vertex in every cell. If a filled column contained no selected $T$,
its bits would be nonincreasing from top to bottom, although the
boundaries require zero at the top and one at the bottom. Each
filled column must therefore select a $T$. Since $T$ occurs only at an incidence and has right horizontal type, horizontal monotonicity forces its row to end in a right-type state. The rows read out at the right boundary form a cover $C(I)\subseteq[m]$.

Write $W_{\mathrm{SC}}=m+1$. For a retained vertex $v$ in
cell $(i,j)\in[m]\times[r]$, set
\[
\omega(v)=
\begin{cases}
W_{\mathrm{SC}}+1,& j=r\text{ and }v\text{ is left-type},\\
W_{\mathrm{SC}},&\text{otherwise}.
\end{cases}
\]
Here $[m]$ indexes the sets and $[r]$ indexes the universe elements.

\begin{figure*}[t]
\centering\includegraphics[width=\textwidth]{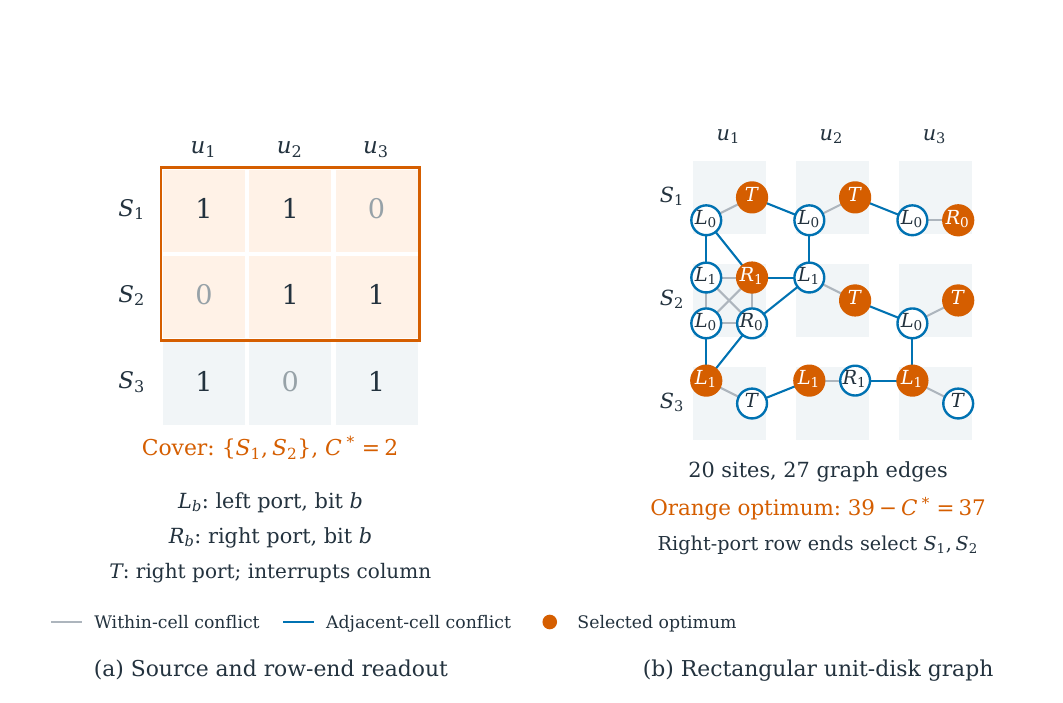}
\caption{The rectangular covering extension.
(a) The incidence matrix for $S_1=\{u_1,u_2\}$,
$S_2=\{u_2,u_3\}$, and $S_3=\{u_1,u_3\}$ has minimum cover size
$C^*=2$. (b) The exact target graph has 20 sites and 27 edges.
States $L_b$ and $R_b$ carry left and right horizontal ports; the
incidence state $T$ carries a right port and interrupts vertical
propagation. Orange sites form an optimum of weight $39-C^*=37$.
Its right-port row ends decode to $\{S_1,S_2\}$.}
\label{fig:setcover}
\end{figure*}

\begin{proposition}[Set-cover encoding]
\label{cor:setcover}
For the preprocessed instance with $m\ge2$ and $r\ge1$, there is a positive-weight unit-disk graph with offset $C_{\mathrm{SC}}=mr(m+1)+m$ such that
\begin{equation*}
\OPT_{\MWIS}(H_{\mathrm{SC}},\omega)=C_{\mathrm{SC}}-C^*.
\end{equation*}
Every target optimum is filled and its row-end readout is a minimum cover.
The target graph uses at most $(4m-4)r$ sites.
\end{proposition}
Figure~\ref{fig:setcover} illustrates the cell states
and row-end readout. The explicit coordinates,
boundary-pruning rules, exact site count, and proof are provided in
Appendix~\ref{app:setcover}. 

\section{Resource requirements}
\label{sec:resources}

\subsection{Complete templates and activity pruning}
For the same $n\ge2$ residual variables, our complete layout and the
fixed triangular layout in the \texttt{map\_qubo} frontend of
\texttt{UnitDiskMapping.jl}~\cite{UnitDiskMappingSnapshot} use
\[
N_{\mathrm{full}}=2n^2-n,\qquad
N_{\mathrm{ref,Q}}=4n^2-2n-4
\]
array sites before relation-induced state deletion. The reference has
four assignment sites and four peripheral sites at each crossing,
plus two sites at each of
the $n-2$ internal turns. The leading quadratic coefficient is
halved. For example, a budget of 400 sites accommodates 14 variables
(378 sites) in our complete template and 10 variables (376 sites)
in the fixed reference template. For MWIS with $m$ source edges,
the respective counts are $2n^2-n-m$ and $4n^2-2n-4-m$ before any
further simplification. The reference MWIS frontend is
\texttt{map\_simple\_wmis}.

For an activity-pruned layout, let $P$ be the sum of the lengths
of the active segments, and let $s_{\mathcal R}$ count states
removed by the pairwise relations:
\[
P=\sum_i(R_i-i)+\sum_i(i-L_i),\qquad
s_{\mathcal R}=\sum_{(i,j)\in\mathcal P}(4-|R_{ij}|).
\]
Every constrained pair belongs to $F$, so its cell has two
active bits and retains all four states before relation deletion.
The sum $s_{\mathcal R}$ therefore counts exactly the removed
vertices. For MWIS, $s_{\mathcal R}=m$. Each cell with $s$ active bits
contributes $s$ to $P$ and $2s$ sites for $s=1,2$. Consequently,
\begin{equation}
N_{\mathrm{clip}}=n+2K_1+4K_2-s_{\mathcal R}
=n+2P-s_{\mathcal R}.
\label{eq:pruned-count}
\end{equation}
In the complete layout $P=n(n-1)$, so the count after relation
deletions is $N_{\mathrm{full}}-s_{\mathcal R}$. In the pruned
layout, for fixed $n$ and relations, minimizing $P$ over variable
orders is equivalent to minimizing the site count.

\subsection{Comparisons on fixed MWIS instances}
\label{sec:resource-benchmarks}

\begin{figure*}[tbp]
\centering
\includegraphics[width=\textwidth]{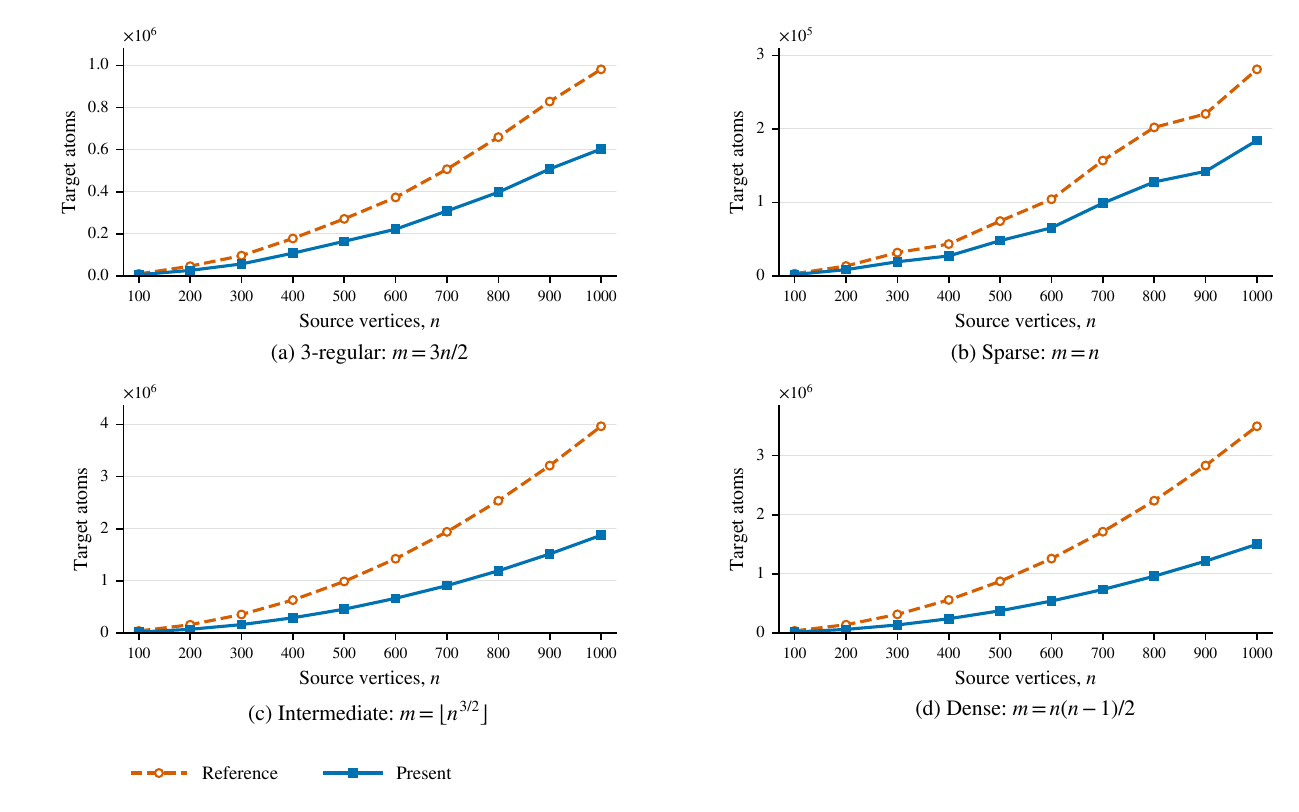}
\caption{Site counts under the shared-order protocol. The four
unit-weight MWIS families are random $3$-regular graphs, sparse graphs
with $m=n$, graphs with $m=\lfloor n^{3/2}\rfloor$, and complete
graphs. Each point uses one fixed instance and the same source
order for the present layout and Weighted. Reference is the
smaller of the simplified Weighted count and the Simple count.
Source preprocessing is disabled.}
\label{fig:reference-present-scaling}
\end{figure*}

The reference uses two frontends from the same pinned release of
\texttt{UnitDiskMapping.jl}~\cite{UnitDiskMappingSnapshot}. Weighted uses \texttt{map\_graph} in weighted mode, reuses
copy-line slots, and applies local simplification. Simple is the
\texttt{map\_simple\_wmis} frontend on the fixed complete template. The reference count is the smaller of their site counts,
and the reduction is reported as a fraction of this reference count.

\paragraph{Shared-order comparison.}
We use 40 unit-weight instances across $n=100,200,\ldots,1000$.
On each graph we select the order $\pi_*$ that minimizes the
present propagation span among seven candidates. At this same
order the reference count is the smaller of the simplified Weighted
mapping and the fixed Simple count. We use the pinned frontend
version and counting protocol in Appendix~\ref{app:resources}.
Figure~\ref{fig:reference-present-scaling} shows this comparison.
At $n=1000$, the present site counts are smaller by $38.6\%$,
$34.4\%$, $52.7\%$, and $57.1\%$ for the four families, respectively.
Across the 40 fixed instances the reduction ranges from
$34.4\%$ to $57.1\%$ under this protocol.

In the $n=m=1000$ sparse instance, choosing $\pi_*$ reduces the
present count from $948{,}218$ to $184{,}422$. At $\pi_*$ the
reference Weighted count falls from $286{,}664$ to $281{,}002$
under local simplification, and the fixed Simple count is
$3{,}996{,}996$. The present count is therefore $34.37\%$ smaller
than the selected reference at this order.

\paragraph{Separate order selection.}
The Weighted frontend can prefer a different order even when the
present span is identical. In the archived sparse $n=100$ instance,
RCM and its reversal both give a present count of $2{,}062$.
The reference count falls from $3{,}166$ to $2{,}384$ under the
reversal, changing the reduction from $34.87\%$ to $13.51\%$.
We therefore evaluate 18 additional paired instances, with three new
seeds in each random family at each of $n=100,200$. Both methods use
the same candidates. The present method minimizes $P$, and the
reference takes the minimum of the separately optimized Weighted count
and the fixed Simple count. Under separate selection the present
layout saves $12.1$--$54.0\%$ of the reference sites across the
18 graphs. The largest decreases relative to the shared-order
percentages occur on the sparse $m=n$ instances.
Table~\ref{tab:paired-ordering} reports the range within each
family and size. These are finite-candidate measurements on
preprocessing-free, unit-weight MWIS inputs. The complete graphs
serve as dense layout controls.

\begin{table*}[t]
\centering
\caption{Paired reductions in site count, in percent, for shared and separately selected orders. Entries give the median [minimum, maximum] over three seeds for each family and size. In both columns the reference is the smaller of the simplified Weighted count and the fixed Simple count.}
\label{tab:paired-ordering}
\begin{tabular}{lr@{\qquad}rr}
\toprule
Family & $n$ & Shared order & Separate order selection \\
\midrule
$3$-regular & 100 & \mbox{42.0 [41.3, 42.1]} & \mbox{33.0 [31.4, 35.9]} \\
$3$-regular & 200 & \mbox{41.1 [40.2, 41.3]} & \mbox{33.2 [32.1, 33.4]} \\
$m=n$ & 100 & \mbox{35.3 [34.6, 36.7]} & \mbox{18.0 [15.9, 23.1]} \\
$m=n$ & 200 & \mbox{34.4 [29.7, 34.5]} & \mbox{14.7 [12.1, 16.4]} \\
$m=\lfloor n^{3/2}\rfloor$ & 100 & \mbox{55.6 [55.4, 56.5]} & \mbox{53.3 [52.9, 53.5]} \\
$m=\lfloor n^{3/2}\rfloor$ & 200 & \mbox{55.8 [55.6, 55.8]} & \mbox{53.8 [53.7, 54.0]} \\
\bottomrule
\end{tabular}
\end{table*}

\paragraph{Benchmark and verification workflow.}
AI-assisted coding supported the benchmark scripts and their check
routines. Appendix~\ref{app:resources} specifies the deterministic
counting protocol and its native Julia comparison. The reproduction
archive contains the fixed inputs, the ported and native count
records, and scripts for rerunning the finite enumerations.
 
\section{Conclusion}
\label{sec:discussion}

One-sided propagation is sufficient for exact MWIS decoding because
the source weights reside on the diagonal and a selected source
vertex propagates a one to every required crossing. Quadratic
terms require consistent active copies, which the endpoint weights
enforce at optimality. These two guarantees use the same four-state
geometry. The rectangular set-cover construction uses the same local
implication with a state that interrupts vertical propagation.

The complete layout halves the leading site-count coefficient of
the fixed complete-grid reference. For the pruned layout, the total
propagation span gives the exact dependence on the variable order,
and the paired benchmarks quantify the effect of allowing the
reference to choose its own order. The geometric cutoff graph has a
positive distance margin. The ten-site example shows that finite
interactions can still change the optimal readout under uniform
detuning rescaling. Extending the physical guarantee requires
analyzing detuning choices and the full interaction energies.

\section*{Acknowledgments}
This work was supported in part by the Quantum Science and Technology-National Science and Technology Major Project under Grant No. 2024ZD0300500, the National Natural Science Foundation of China Grants No. 92465202, 12501450, and the China Postdoctoral Science Foundation under Grant Number No. 2026T190829.

\section*{Data and code availability}
The accompanying reproduction archive contains the input graphs,
permutations, site counts, and verification scripts used for the
figures, tables, and worked examples. The associated code repository
also provides constructions and verification procedures for additional
reductions (\url{https://github.com/qhub-cn/rydkit.git}).

\section*{AI Use Disclosure}
AI-assisted tools supported the manuscript and its reproduction
materials. ChatGPT (OpenAI Codex) assisted in revising the
organization and exposition of the proofs, checking the wording
against the exact constructions and benchmark data, and preparing
the figures and verification scripts. The supporting archive
includes explicit constructions, finite enumerations, and
independent count checks, including the worked cases in Appendix~\ref{app:examples}.
\appendix

\section{Proof of Lemma~\ref{lem:geometry}}
\label{app:geometry}

\begin{proof}
Use the coordinates in Sec.~\ref{sec:four-state-geometry}.
All positions are distinct: different states within a cell
have different coordinates, while a change in either cell
index produces a coordinate displacement of at least
$27/20-3/5=3/4$.

Within each cell,  the maximum squared distance is
\[
2\left(\frac35\right)^2=\frac{18}{25}<1.
\]
Thus every cell forms a clique.

Consider horizontally adjacent cells $(i,j)$ and
$(i,j+1)$. The horizontal separation between
$v_{ij}^{ab}$ and $v_{i,j+1}^{a'b'}$ is
\[
\left|\frac{27}{20}+\frac35(a'-a)\right|.
\]
When $(a,a')=(1,0)$, this separation equals $3/4$,
while the vertical separation is at most $3/5$.
Consequently,
\[
\left\|
q(v_{ij}^{ab})-q(v_{i,j+1}^{a'b'})
\right\|^2
\le
\left(\frac34\right)^2+
\left(\frac35\right)^2
=\frac{369}{400}<1.
\]
For every other pair $(a,a')$, the horizontal
separation is at least $27/20>1$.
Hence the horizontal adjacency rule holds exactly.

For $(i,j)$ and $(i+1,j)$, the vertical separation is
$\lvert-27/20+3(b'-b)/5\rvert$. It equals $3/4$ for
$(b,b')=(0,1)$ and is at least $27/20$ otherwise.
The horizontal separation is at most $3/5$, so the
same edge bound applies. Both rules cover interfaces
involving the diagonal vertices.

It remains to exclude edges between nonadjacent cells.
If the cell indices differ by at least two along either
coordinate, the corresponding separation is at least
\[
2\cdot\frac{27}{20}-\frac35
=\frac{21}{10}>1.
\]
Otherwise, nonadjacent cells must differ by one index
in both directions. Their coordinate separations are
then at least $3/4$ in each direction, so
\[
\|q(u)-q(v)\|^2
\ge
2\left(\frac34\right)^2
=\frac98>1.
\] 

The construction therefore realizes exactly the
adjacency rules stated in Lemma~\ref{lem:geometry}.
Moreover, every edge has length at most
$\sqrt{369}/20$, and every nonedge has distance
at least $\sqrt{9/8}$.
Hence the required constant separation margin is
\[
\eta=
\min\left\{
1-\frac{\sqrt{369}}{20},
\sqrt{\frac98}-1
\right\}>0,
\]
independent of $n$.
\end{proof}
 
\section{Canonical encodings in both layouts}
\label{app:pruning}
Each required pair $(i,j)\in F$ satisfies $j\le R_i$ and $i\ge L_j$, so both coordinates remain active there. In the complete layout all copies are active. In the pruned layout every cell between a diagonal and an active endpoint is retained. For a filled independent set $I$, let $(a_{ij},b_{ij})$ be the
state of its unique vertex in each retained cell and write
$x=x(I)$. Then
\begin{align*}
x_i&\le a_{i,i+1}\le\cdots\le a_{i,R_i},&&R_i>i,\\
x_i&\le b_{i-1,i}\le\cdots\le b_{L_i,i},&&L_i<i.
\end{align*}
Canonical choices make each active segment constant. If a retained cell lies outside that segment, its corresponding bit equals one and cannot introduce an outward decrease. Allowed relations are respected at interaction cells by source feasibility. No other intercell conflicts exist, proving that $I_x$ is independent exactly when its requested constrained states are retained.

\section{Weight bounds and consistency for quadratic objectives}
\label{app:quadratic-proof}

\subsection{Positive weights and coefficient bounds}

At most two endpoint deductions affect a state vertex. Its weight is
therefore at least $\Lambda-|J_{ij}|\ge1$ and at most
$3\Lambda+|J_{ij}|\le4A+3$. Since $1\le\kappa_i\le2$, a diagonal
weight lies between $\Lambda-h_i\ge1$ and
$2\Lambda+|h_i|\le3A+2$. This proves Eq.~\eqref{eq:weight-range}.

For a target independent set $I$, let $z(I)$ count empty retained cells.
For each nonempty segment $s$ of variable $i(s)$ in the chosen layout, denote its terminal bit
by $\beta_s(I)$, taking zero when the endpoint cell is empty. Define
\begin{align}
T(I)&=\sum_{s\text{ active}}
\bigl(x_{i(s)}(I)-\beta_s(I)\bigr),\\
\Phi(I)&=\sum_i h_ix_i(I)+
\sum_{i<j:\,v_{ij}^{11}\in I}J_{ij}.
\label{eq:imbalance}
\end{align}
Grouping the vertex weights gives
\begin{equation}
\omega(I)=\CQ-z(I)\dQ+\Lambda T(I)-\Phi(I).
\label{eq:weight-identity}
\end{equation}
Each coefficient in $\Phi(I)$ is multiplied by a binary indicator, so
\begin{equation}
-A_-\le\Phi(I)\le A_+.
\label{eq:objective-bounds}
\end{equation}

\subsection{Propagation deficits and canonical separation}

\begin{lemma}[Propagation deficit]
\label{lem:deficit}
Every target independent set satisfies $T(I)\le2z(I)$.
Filled sets satisfy $T(I)\le0$, and filled noncanonical sets
satisfy $T(I)\le-1$.
\end{lemma}

\begin{proof}
A positive segment contribution in $T(I)$ requires diagonal bit one and endpoint
bit zero. A segment with no empty cell cannot have this transition by
Lemma~\ref{lem:geometry}. Assign each positive contribution to the first
empty cell on that segment. Each empty cell lies on at most one
horizontal and one vertical nonempty segment, so the total positive
contribution is at most $2z(I)$.

If every cell is filled, all segment contributions are nonpositive.
If such a set is noncanonical, an active copy is one while its diagonal
bit is zero. The copy lies on a contiguous occupied segment ending at
the site of the endpoint deduction. This is the outer cell of that
path for the complete layout and the last active cell for the pruned
layout. Monotonicity carries the one to this terminal bit, giving a
contribution of $-1$. In the pruned layout the fixed inactive bits lie
outside the compensated segments.
\end{proof}
\subsection{Proof of Lemma~\ref{lem:canonical-separation}}
\begin{proof} 
For a nonfilled independent set $I$, we have $z(I)\ge1$.
By Lemma~\ref{lem:deficit} and
Eqs.~\eqref{eq:weight-identity}--\eqref{eq:objective-bounds},
\begin{equation*}
\begin{aligned}
\omega(I)
&\le\CQ-z(I)\dQ+2z(I)\Lambda+A_-\\
&=\CQ-z(I)\Lambda+A_-\\
&\le\CQ-\Lambda+A_-\\
&=\CQ-A_+-1.
\end{aligned}
\end{equation*}
For a filled noncanonical independent set,
$z(I)=0$ and $T(I)\le-1$.
The same weight bound follows from
Eqs.~\eqref{eq:weight-identity} and
\eqref{eq:objective-bounds}.

For every feasible assignment $x\in\cD$,
the canonical encoding $I_x$ is independent
by Appendix~\ref{app:pruning}.
It satisfies
\[
r(I_x)=0,\qquad
T(I_x)=0,\qquad
\Phi(I_x)=Q(x)-c.
\]
Substituting these identities into
Eq.~\eqref{eq:weight-identity} yields
Eq.~\eqref{eq:canonical-identity}.

Conversely, a canonical independent set selects
the state corresponding to its diagonal readout
at every constrained pair.
Since these states are retained, the readout
satisfies every prescribed relation and
therefore belongs to $\cD$.
\end{proof}

\section{Worked quadratic examples}
\label{app:examples}

\subsection{A constrained quadratic instance}

Consider
\begin{equation}
\begin{aligned}
Q(x)&=-3x_1-4x_2-x_3+2x_1x_2-2x_2x_3,\\
x_1+x_2&\le1.
\end{aligned}
\label{eq:ablation}
\end{equation}
Here $A=12$, $\Lambda=13$, and $\dQ=39$.
Activity pruning retains $K=2$ cells, with 10 vertices
and 17 edges. The retained endpoints are
$R=(2,3,3)$ and $L=(1,1,2)$, giving $\CQ=78$.

The diagonal weights are $(16,30,14)$.
Cell $(1,2)$ retains states $00,01,10$ with weights
$39,26,26$, respectively, while state $11$ is deleted
to enforce the mutual-exclusion constraint.
Cell $(2,3)$ retains all four states $00,01,10,11$
with weights $39,26,26,15$, respectively.

The source problem has a unique optimum $x^*=011$,
with $Q(x^*)=-7$.
Solving the constructed MWIS instance yields the unique
optimal independent set
\[
I^*=\{d_2,d_3,v_{12}^{01},v_{23}^{11}\},
\]
with weight
\[
\omega(I^*)=30+14+26+15=85
=\CQ-Q(x^*).
\]
Reading the diagonal vertices gives $x=011$,
which agrees with the source optimum.

\subsection{An infeasible quadratic instance}

Consider
\[
Q(x)=-10x_2,\qquad
R_{12}=\{00,01\},\quad
R_{13}=\{10,11\}.
\]
The two relations require $x_1=0$ and $x_1=1$,
respectively, so the source instance is infeasible.

Here $A=10$, $A_+=0$, $\Lambda=11$, and $\CQ=99$.
The constructed graph has 9 vertices and 8 edges.
Solving the MWIS instance gives an optimal weight of 98.
One optimal independent set is
\[
I^*=\{d_2,v_{12}^{01},v_{13}^{10},v_{23}^{10}\},
\]
with
\[
\omega(I^*)=21+22+22+33=98.
\]
Its diagonal readout is $x=010$, which violates
the relation $R_{13}$.
Moreover,
\[
\OPT_{\MWIS}(H,\omega)=98<\CQ-A_+=99,
\]
correctly identifying the source instance as infeasible.

\subsection{An instance with two optimal solutions}

Consider
\[
Q(x)=-x_1-2x_2-x_3+2x_1x_3,
\]
subject to
\[
R_{12}=R_{13}=R_{23}=\{01,10,11\}.
\]
These relations forbid the assignment $00$ to every
variable pair. The feasible assignments are
$011$, $101$, $110$, and $111$, with
\[
Q(011)=Q(110)=-3,\quad
Q(101)=0,\quad Q(111)=-2.
\]
Thus the source problem has exactly two optimal
solutions, $011$ and $110$.

Here $A=6$, $\Lambda=7$, $\dQ=21$, and $\CQ=63$.
The constructed graph has 12 vertices and 17 edges.
The diagonal weights are $(8,16,8)$. In state order $01,10,11$,
the weights of cells $(1,2)$, $(1,3)$, and $(2,3)$ are
$(14,21,14)$, $(14,14,5)$, and $(21,14,14)$, respectively.
Solving the MWIS instance yields exactly two optimal
independent sets:
\begin{align*}
I_{011}
&=\{d_2,d_3,v_{12}^{01},v_{13}^{01},v_{23}^{11}\},\\
I_{110}
&=\{d_1,d_2,v_{12}^{11},v_{13}^{10},v_{23}^{10}\}.
\end{align*}
Both have weight
\[
\omega(I_{011})=\omega(I_{110})
=8+16+3(14)=66
=\CQ-(-3).
\]
Their diagonal readouts are $011$ and $110$,
respectively, agreeing with the two source optima.

\subsection{Complete-layout endpoint check}
\label{app:complete-endpoints}

Let $Q(x)=x_1+x_2+x_3+x_1x_2+x_2x_3$ with no additional relation.
The unique source optimum is $000$. Here $\Lambda=6$ and
$\dQ=18$. The complete graph has 15 sites and $K=3$.
With deductions at its true endpoints the unique optimal
configuration has weight $54$ and selects $00$ in all three cells.
If the complete graph is instead weighted using the activity-pruned
endpoints $R=(2,3,3)$ and $L=(1,1,2)$, the maximum remains $54$,
but the unused cell $(1,3)$ can take any of its four states.
The diagonal readout is still $000$ for all four maxima. This
enumeration checks the role of layout-specific endpoint deductions.

\section{Proof of Proposition~\ref{cor:setcover}}
\label{app:setcover}

\begin{proof}
Let $M_{ij}=1$ if $u_j\in S_i$, and $M_{ij}=0$ otherwise.
Work in coordinates scaled by 20, so the squared-distance
cutoff is $400$. Use cell origins
$B_{ij}=(27(j-1),-27(i-1))$.
For each nonincidence cell, place
\[
L_b=B_{ij}+(0,12b),\qquad
R_b=B_{ij}+(12,12b),
\quad b\in\{0,1\}.
\]
At an incidence cell, replace $R_0,R_1$ by
$T=B_{ij}+(12,6)$, retaining $L_0$ and $L_1$.

Every cell forms a clique. Horizontal conflicts join
right-type states in one cell to left-type states in
the next, forbidding right-to-left transitions.
Vertical conflicts join bit zero in an upper cell
to bit one in the cell below.
The state $T$ has no vertical intercell conflicts,
since its vertical separation from neighboring cells
is at least $21>20$.
All other distances satisfy the corresponding edge
and nonedge bounds established in
Appendix~\ref{app:geometry}.
Thus, the construction realizes the required unit-disk
graph.

For each column $j$, define
\[
f_j=\min\{i:M_{ij}=1\},\qquad
\ell_j=\max\{i:M_{ij}=1\}.
\]
Delete bit-one states in rows $i\le f_j$ and
bit-zero states in rows $i\ge\ell_j$, retaining $T$.
If $f_j=\ell_j$, the corresponding incidence cell
contains only $T$.
Every cell remains nonempty.

Recall that $W_{\mathrm{SC}}=m+1$, with an additional unit reward
for left-type states in the last column.
Given a source cover $C$, choose right-type states
throughout each row $i\in C$ and left-type states
throughout every other row.
For each element $u_j$, choose a witness row
$i\in C$ with $u_j\in S_i$.
Select $T$ at every incidence cell in a row of $C$, including the witness, using bit zero above the witness
and bit one below it wherever a vertical bit is
required.
These choices satisfy all boundary restrictions
and intercell conflicts, yielding a filled
independent set $I_C$ of weight
\[
\omega(I_C)=mrW_{\mathrm{SC}}+m-|C|.
\]

Conversely, consider any filled independent set $I$.
If a column contained no selected $T$, its vertical
bits would form a nonincreasing sequence from top
to bottom.
However, the boundary restrictions require bit zero
at the top and bit one at the bottom, a contradiction.
Thus, every column contains a selected $T$.
Since $T$ occurs only at incidence cells and its
right horizontal type propagates to the row end,
the rows with right-type states in the last column
form a cover $C(I)$.
Its weight is
\[
\omega(I)=mrW_{\mathrm{SC}}+m-|C(I)|.
\]

A nonfilled independent set has at most $mr-1$
occupied cells and therefore satisfies
\[
\omega(I)\le(mr-1)W_{\mathrm{SC}}+m<mrW_{\mathrm{SC}}.
\]
Since choosing all sets yields a filled independent
set of weight $mrW_{\mathrm{SC}}$, every target optimum is filled.
Consequently,
\[
\OPT_{\MWIS}(H_{\mathrm{SC}},\omega)
=
mrW_{\mathrm{SC}}+m-\min_{C\text{ cover}}|C|.
\]
Moreover, every target optimum decodes to a
minimum set cover.

It remains to establish the vertex count.
Let $\nu_j$ be the number of incidences in column $j$.
Before boundary pruning, the column contains
$4m-\nu_j$ vertices.
The boundary restrictions remove
$2(f_j-1)+2(m-\ell_j)+2$ vertices, giving
\[
N_j=2m-\nu_j+2(\ell_j-f_j).
\]
For $\nu_j=1$, we have
$N_j=2m-1\le4m-4$ when $m\ge2$.
For $\nu_j\ge2$,
\[
N_j\le2m-2+2(m-1)=4m-4.
\]
Writing $\nu=\sum_i|S_i|$, the exact site count is
\[
N_{\mathrm{SC}}=2mr-\nu+2\sum_{j=1}^r(\ell_j-f_j)
\le(4m-4)r.
\]
The bound is attained within the rectangular
layout when every column has incidences only
in the first and last rows.
\end{proof}
 
For positive costs $\gamma_i>0$, replace $W_{\mathrm{SC}}$ by
$\sum_i \gamma_i+1$ and assign weight $W_{\mathrm{SC}}+\gamma_i$
to left-type states in the last column of row $i$.
The same argument gives
\[
\OPT_{\MWIS}
=
mrW_{\mathrm{SC}}+\sum_i \gamma_i
-\min_{C\text{ cover}}\sum_{i\in C}\gamma_i.
\]
The coordinates and vertex count remain unchanged.

\section{Finite interactions in the ten-site instance}
\label{app:finite-interaction}

At zero drive, put $\mathbf r_v=R_bq(v)$ and
$\Delta_v=\lambda\omega(v)$, and measure energy in units of
$C_6/R_b^6$. With $t=\lambda/(C_6/R_b^6)$, a configuration $S$ has energy
\[
\begin{aligned}
e_t(S)&=U(S)-t\omega(S),\\
U(S)&=\sum_{\{u,v\}\subseteq S}\|q(u)-q(v)\|_2^{-6}.
\end{aligned}
\]
Here $S\subseteq V(H)$ ranges over all atom occupations, including simultaneous
excitation on graph edges. In the ten-site example of
Appendix~\ref{app:examples}, write $u^{ab}=v_{12}^{ab}$ and
$v^{ab}=v_{23}^{ab}$. The graph optimum
$I^*=\{d_2,d_3,u^{01},v^{11}\}$ has energy
$0.453582724-85t$. The independent competing configuration
$\{d_2,u^{01},v^{10}\}$ has energy $0.194656780-82t$, and the
edge-violating configuration $\{d_1,d_2,u^{00},v^{10}\}$ has
energy $2.670790186-111t$. Pairwise comparison imposes the
incompatible bounds $t\ge0.086308648\ldots$ and
$t\le0.085277210\ldots$. The coefficients shown here are rounded.

To check the diagonal readout over the entire positive-$t$ range,
we compute $U$ exactly as a rational for each of the $2^{10}=1024$
occupations, using the integer squared distances of the coordinates
scaled by 20. For each of the 128 occupations $S$ that read out
$011$, we intersect $t>0$ with the half-lines
\[
t\,[\omega(S)-\omega(S')]\ \ge\ U(S)-U(S')
\]
for every $S'$. All 128 intersections are empty, including the
possibility of tied ground states. This computation uses fixed
coordinates, fixed relative detunings, and zero drive.

\section{Resource counts}
\label{app:resources}

\subsection{Source instances}

Figure~\ref{fig:reference-present-scaling} and
Table~\ref{tab:mwis-scaling-all-points} use the same 40 MWIS instances. For each $n=100,200,\ldots,1000$, we construct four simple undirected
graphs on $n$ vertices: a random $3$-regular graph, a sparse graph with
$m=n$ edges, a graph with $m=\lfloor n^{3/2}\rfloor$ edges, and the
complete graph $K_n$ with $m=n(n-1)/2$ edges. Every source vertex has
unit weight. We count one atom per site in the target unit-disk graph.

We generate each random $3$-regular graph using the NetworkX function
\texttt{random\_regular\_graph}. Since every vertex has degree three,
the graph has $m=3n/2$ edges. For the two families with $m=n$ and
$m=\lfloor n^{3/2}\rfloor$, we first draw a uniformly random permutation
$(v_1,\ldots,v_n)$ of the vertices. The edges $\{v_{2j-1},v_{2j}\}$
for $j=1,\ldots,n/2$ form a perfect matching ensuring that every vertex has at least one neighbor. We then sample $m-n/2$
additional edges uniformly without replacement from the remaining
unordered vertex pairs.   

For a graph of size $n$, we use the random seed $20261124+10^5 k+n$,
where $k=1,2,3$ indexes the $3$-regular, $m=n$, and
$m=\lfloor n^{3/2}\rfloor$ families, respectively. The fourth family
consists of the complete graphs $K_n$, each with all $\binom{n}{2}$
possible edges.

\subsection{Present construction and variable ordering}

Let $\pi=(v_1,\ldots,v_n)$ give the source vertices in layout order.
Let $N_G[v_i]$ denote the closed neighborhood of $v_i$ in $G$,
and write $B_i=\{j:v_j\in N_G[v_i]\}$ for its positions. Then
$L_i=\min B_i$ and $R_i=\max B_i$. Variable $v_i$ propagates
horizontally from position $i$ to $R_i$ and vertically from $i$ to
$L_i$. Summing these segment lengths gives the propagation span
\begin{equation}
  P(\pi)=\sum_{i=1}^{n}\big[(R_i-i)+(i-L_i)\big].
  \label{eq:app-mwis-benchmark-span}
\end{equation}

The layout retains one diagonal atom per source vertex. In an
off-diagonal cell $(i,j)$ with $i<j$, the horizontal bit is active if
$j\le R_i$ and the vertical bit is active if $i\ge L_j$. We omit a
cell with no active bits. A cell with one active bit retains two states,
with the inactive bit fixed to one. A cell with two active bits contains
four states before enforcing the MWIS constraint. Each source edge has
both bits active and excludes precisely the state $(1,1)$. Counting two
sites per active bit and subtracting one per source edge therefore gives
\begin{equation}
  N_{\mathrm{clip}}(\pi)=n+2P(\pi)-m.
  \label{eq:app-mwis-benchmark-count}
\end{equation}
For $K_n$, every order gives $P(\pi)=n(n-1)$ and hence
$N_{\mathrm{clip}}=(3n^2-n)/2$.

Since $n$ and $m$ are fixed for each instance, minimizing $P(\pi)$
also minimizes the present atom count over the candidate orders.
In generation order, the seven candidates are
\texttt{input} (increasing source label), \texttt{reverse},
\texttt{rcm} (reverse Cuthill--McKee), \texttt{rcm\_reverse}
(the reverse of that permutation), \texttt{degree\_asc},
\texttt{degree\_desc}, and \texttt{hub\_middle}. The RCM candidate uses
SciPy on the symmetric source adjacency matrix. Both degree sorts break
ties by increasing source label. For \texttt{hub\_middle}, we place
vertices in descending degree order, starting at the center and moving
outward, filling the left position first when two positions are equally
far from the center.

For duplicate permutations we keep the first name in the sequence above.
We then choose the smallest span, breaking ties between distinct
permutations by lexicographic order of their retained names. Denote the
result by $\pi_*$. This rule selected \texttt{rcm} for all 30 random
instances and \texttt{hub\_middle} for all ten complete graphs, for which
every permutation has the same span. The reported Present count is
$N_{\mathrm{clip}}(\pi_*)$.

\subsection{Reference constructions and reported ratio}

We compare the present construction with two MWIS mappings implemented
in \texttt{UnitDiskMapping.jl}, version~0.5.1
(commit \texttt{86cb5f0e})~\cite{Nguyen2023}.
In the shared-order protocol the weighted mapping uses $\pi_*$.
The weighted mapping introduces vertices in
this order and assigns each copy line the first available horizontal
slot. Once a vertex and all its neighbors have been introduced, its
slot becomes available for reuse. The mapping then replaces local
crossings and turns with the prescribed gadgets.

We denote the number of array sites after gadget replacement by
$N_{\mathrm{W,before}}$ and the count after local simplification by
$N_{\mathrm{W,after}}$. Both counts use the same order $\pi_*$. Each
application of the default simplification rule removes two sites.
We use the default limit of ten passes and stop earlier if a full
pass leaves the layout unchanged.

For the 30 random instances, we obtain both weighted counts using a
Python port of the specified Julia source. For the complete graphs
$K_n$, counting the sites in the weighted templates gives
$N_{\mathrm{W,before}}=(7n^2-9n+2)/2$ for $n\geq3$. Exactly two
simplifications apply at the ends of the layout, removing four sites
in total and giving $N_{\mathrm{W,after}}=(7n^2-9n-6)/2$. We verified
both formulas against the Python port for $n=3,\ldots,32$ and $n=100$.

The second mapping, \texttt{map\_simple\_wmis}, uses a complete
triangular grid. Each source edge contributes seven sites, each
nonedge contributes eight, and each of the $n-2$ internal turns
contributes two. The total therefore depends only on $n$ and $m$:
\begin{align}
  N_{\mathrm{simple}}
  &= 7m + 8\left[\binom{n}{2}-m\right] + 2(n-2) \nonumber\\
  &= 4n^2-2n-4-m, \qquad n\geq2.
  \label{eq:app-mwis-benchmark-simple}
\end{align}
This mapping encodes source rewards as small perturbations to the
target weights. We therefore multiply each unit reward by
$\varepsilon_{\mathrm{pert}}=1/[16(n+1)]$, giving a total perturbation of
$n\varepsilon_{\mathrm{pert}}<1/16$. This uniform positive rescaling preserves the
optimal source independent sets and leaves the array-site count
unchanged.

For each instance, we take the smaller of the two reference counts
and report the ratio of the present count to that reference:
\begin{align}
  N_{\mathrm{ref}}
  &= \min\{N_{\mathrm{W,after}},N_{\mathrm{simple}}\}, \nonumber\\
  \rho
  &= \frac{N_{\mathrm{clip}}(\pi_*)}{N_{\mathrm{ref}}}.
\end{align}
The complete-grid mapping gives the smaller count for the nine
instances with $m=\lfloor n^{3/2}\rfloor$ and $n\geq200$. The weighted
mapping gives the smaller count for the remaining 31 instances.

The reproduction archive records both endpoint spans and both
Weighted stages for the 40 instances, including the sparse example
discussed in Sec.~\ref{sec:resource-benchmarks}.

\subsection{Separately selected orders}
\label{app:separate-orders}
We apply the same generators, candidate rules, and ten-pass Weighted
simplification to 18 additional fixed inputs. For each family index
$k=1,2,3$, each $n\in\{100,200\}$, and each replicate $b\in\{1,2,3\}$,
we use seed $20261124+10^5 k+10^3b+n$. All source weights are one,
and source preprocessing is disabled. On each graph, let $\Pi$ be the
deduplicated candidate set. We compare
\[
\min_{\pi\in\Pi}N_{\mathrm{clip}}(\pi)
\quad\text{with}\quad
\min\!\left\{N_{\mathrm{simple}},
             \min_{\pi\in\Pi}N_{\mathrm{W,after}}(\pi)\right\}.
\]
The names break count ties lexicographically. The accompanying
data archive fixes the source graphs and every candidate
permutation, and records the counts before and after simplification.

\paragraph{Native implementation check.}
We ran the pinned Julia implementation with Julia 1.11.7 on six
additional graphs, one in each sampled family at each of $n=12,24$.
For each graph we used a fixed RCM order and its reversal.
All twelve native before/after count pairs agreed with the Python
port under the default ten-pass limit. We also verified the
complete-template formulas using native \texttt{map\_qubo} and
\texttt{map\_simple\_wmis} for $n=2,\ldots,20$.
The reproduction archive includes the inputs and both count stages.

\begin{table*}[tbp]
\centering\small
\caption{Atom counts for the 40 unit-weight MWIS instances plotted in Fig.~\ref{fig:reference-present-scaling}, with source preprocessing disabled. Each point corresponds to one fixed instance. Present uses the selected variable order. Reference is the smaller of the simplified weighted-layout count in the same order and the complete-grid count. The complete-grid count is selected only for the $m=\lfloor n^{3/2}\rfloor$ family at $n\geq 200$. Each ratio is Present divided by Reference, rounded to four decimal places.}
\label{tab:mwis-scaling-all-points}
\begin{tabular}{r@{\qquad}rrr@{\qquad}rrr}
\toprule
& \multicolumn{3}{c}{3-regular ($m=3n/2$)} & \multicolumn{3}{c}{Sparse ($m=n$)} \\
\cmidrule(lr){2-4}\cmidrule(lr){5-7}
$n$ & Present & Reference & Ratio & Present & Reference & Ratio \\
\midrule
100 & 5,888 & 9,958 & 0.5913 & 2,062 & 3,166 & 0.6513 \\
200 & 26,760 & 46,719 & 0.5728 & 8,666 & 13,562 & 0.6390 \\
300 & 57,304 & 96,855 & 0.5916 & 19,398 & 31,785 & 0.6103 \\
400 & 108,462 & 178,239 & 0.6085 & 27,284 & 43,437 & 0.6281 \\
500 & 164,792 & 271,429 & 0.6071 & 48,054 & 74,679 & 0.6435 \\
600 & 222,386 & 373,301 & 0.5957 & 65,562 & 104,349 & 0.6283 \\
700 & 308,908 & 507,237 & 0.6090 & 99,030 & 156,994 & 0.6308 \\
800 & 397,812 & 659,500 & 0.6032 & 127,922 & 202,052 & 0.6331 \\
900 & 508,414 & 828,638 & 0.6136 & 142,322 & 220,409 & 0.6457 \\
1000 & 603,022 & 981,362 & 0.6145 & 184,422 & 281,002 & 0.6563 \\
\midrule
& \multicolumn{3}{c}{$m=\lfloor n^{3/2}\rfloor$} & \multicolumn{3}{c}{Complete ($m=n(n-1)/2$)} \\
\cmidrule(lr){2-4}\cmidrule(lr){5-7}
$n$ & Present & Reference & Ratio & Present & Reference & Ratio \\
\midrule
100 & 16,440 & 37,543 & 0.4379 & 14,950 & 34,547 & 0.4327 \\
200 & 69,566 & 156,768 & 0.4438 & 59,900 & 139,097 & 0.4306 \\
300 & 159,710 & 354,200 & 0.4509 & 134,850 & 313,647 & 0.4299 \\
400 & 290,312 & 631,196 & 0.4599 & 239,800 & 558,197 & 0.4296 \\
500 & 455,644 & 987,816 & 0.4613 & 374,750 & 872,747 & 0.4294 \\
600 & 664,424 & 1,424,100 & 0.4666 & 539,700 & 1,257,297 & 0.4293 \\
700 & 907,794 & 1,940,076 & 0.4679 & 734,650 & 1,711,847 & 0.4292 \\
800 & 1,192,005 & 2,535,769 & 0.4701 & 959,600 & 2,236,397 & 0.4291 \\
900 & 1,515,900 & 3,211,196 & 0.4721 & 1,214,550 & 2,830,947 & 0.4290 \\
1000 & 1,875,034 & 3,966,374 & 0.4727 & 1,499,500 & 3,495,497 & 0.4290 \\
\bottomrule
\end{tabular}
\end{table*}

% REVTeX selects apsrev4-2 and emits the control record for reference titles.
\bibliography{references}
\end{document}